\documentclass[11pt,english,epsf]{scrartcl}
\usepackage{lmodern}
\usepackage[T1]{fontenc}
\usepackage[latin9]{inputenc}
\usepackage{geometry}
\usepackage{amsmath}
\usepackage{amssymb}

\newtheorem{theorem}{Theorem}

\newcommand {\dfn} {\stackrel{\Delta} {=}}
\newcommand {\exe} {\stackrel{\cdot} {=}}
\newcommand {\lexe} {\stackrel{\cdot} {\le}}

\newcommand {\bu} {\mbox{\boldmath $u$}}

\newcommand {\bx} {\mbox{\boldmath $x$}}
\newcommand {\by} {\mbox{\boldmath $y$}}

\newcommand {\bE} {\mbox{\boldmath $E$}}

\newcommand {\bU} {\mbox{\boldmath $U$}}

\newcommand {\bX} {\mbox{\boldmath $X$}}
\newcommand {\bY} {\mbox{\boldmath $Y$}}

\newcommand{\calC}{{\cal C}}

\newcommand{\calE}{{\cal E}}

\newcommand{\calI}{{\cal I}}

\newcommand{\calS}{{\cal S}}
\newcommand{\calT}{{\cal T}}
\newcommand{\calU}{{\cal U}}

\newcommand{\calX}{{\cal X}}
\newcommand{\calY}{{\cal Y}}

\allowdisplaybreaks

\begin{document}
\thispagestyle{empty}
\title{Exact Random Coding Error Exponents of Optimal Bin Index Decoding
\thanks{This research was supported by the Israel Science Foundation (ISF),
grant no.\ 412/12.}
}
\author{Neri Merhav
}
\date{}
\maketitle

\begin{center}
Department of Electrical Engineering \\
Technion - Israel Institute of Technology \\
Technion City, Haifa 32000, ISRAEL \\
E--mail: {\tt merhav@ee.technion.ac.il}\\
\end{center}
\vspace{1.5\baselineskip}
\setlength{\baselineskip}{1.5\baselineskip}

\begin{center}
{\bf Abstract}
\end{center}
\setlength{\baselineskip}{0.5\baselineskip}
We consider ensembles of channel codes that are partitioned into bins, and
focus on analysis of exact random coding error exponents associated
with optimum
decoding of the index of the bin to which the transmitted codeword belongs. 
Two main conclusions arise from this analysis: 
(i) for independent random selection of codewords within a given type
class, the random coding exponent of optimal bin index decoding is 
given by the ordinary random coding exponent function, computed at the rate of
the entire code, independently of the exponential rate of the size of the bin.
(ii) for this ensemble of codes, sub-optimal bin index decoding, that is based
on ordinary maximum
likelihood (ML) decoding, is as good as the optimal bin index decoding in terms of the
random coding error exponent achieved. Finally, for the sake of completeness,
we also outline how our analysis of exact
random coding exponents extends to the hierarchical 
ensemble that correspond to superposition coding and optimal
decoding, where for each bin, first, a cloud center is drawn at random, and then the
codewords of this bin are drawn
conditionally independently given the cloud center. For this ensemble,
conclusions (i) and (ii), mentioned above, no longer hold necessarily in general.

\vspace{0.2cm}


\setlength{\baselineskip}{2\baselineskip}
\newpage

\section{Introduction}

In multiuser information theory, one of the most frequently encountered
building blocks is the notion of
{\it superposition coding}, namely, coding with an hierarchical structure, in which
the codebook is naturally partitioned into {\it bins}, or {\it clouds}. The
original idea of superposition coding
dates back to Cover \cite{Cover72}, who proposed it in the context of broadcast 
channels (see also \cite{Cover98}, \cite[Section 15.6]{CT06} and references therein).
Later on, it has been proved extremely useful in a much wider variety of coded
communication settings,
including the wiretap channel \cite{CK78}, \cite{Wyner75}, the Gel'fand--Pinsker channel
\cite{GP80} (and in duality, Wyner-Ziv source encoding \cite{WZ76}), the relay
channel \cite{CG79}, the
interference channel \cite{Carleial79}, the multiple access channel
\cite{GRUW01}, and channels with feedback \cite{CL81}, \cite{Ozarow84}, just to
name a few.

Generally speaking, the aim of superposition coding is to encode pairs of
messages jointly, such that each message pair is mapped into a single codeword. 
To this end, the codebook is constructed with an hierarchical structure of
bins (or clouds), such that a receiver that operates under relatively good
channel conditions (high SNR) can decode reliably both messages, whereas a receiver
that works under relatively bad channel conditions (low SNR) can decode reliably at least 
one of the messages, the one which consists of the
index of the bin to which the codeword belongs. 

This hierarchical structure of partitioning into bins is applicable even
in achievability schemes with simple code ensembles, where all codewords are
drawn independently under a certain distribution.
Consider, for example, a random code of size $M_1=e^{nR_1}$,
where each codeword $\bx_i=(x_{1,i},x_{2,i},\ldots,x_{n,i})$,
$i=0,1,\ldots,M_1-1$,
is selected independently at random
with a uniform distribution over a given type class.
The code is then divided into $M=e^{nR}$ ($R\le R_1$) bins $\{\calC_w\}_{w=0}^{M-1}$,
$\calC_w=\{\bx_{wM_2},\bx_{wM_2+1},\ldots,\bx_{(w+1)M_2-1}\}$, where
$M_2=M_1/M=e^{n(R_1-R)}\dfn e^{nR_2}$. Assuming that the choice of the index
$i$ of the transmitted codeword is governed by the uniform distribution over
$\{0,1,\ldots,M_1-1\}$, our focus, in this paper, will be on the user that
decodes merely the index $w$ of the bin $\calC_w$ that contains $\bx_i$,
namely, $w=\lfloor i/M_2\rfloor$. 
This problem setting, including the above described random coding ensemble, 
is the very same as the one encountered from the viewpoint of the legitimate
receiver in the achievability scheme of the wiretap channel model
\cite{Wyner75}, as well as the decoder in the direct part of the
Gel'fand--Pinsker channel \cite{GP80}.  

Denoting the channel output vector by $\by=(y_1,\ldots,y_n)$ and the channel
transition probability function by $P(\by|\bx)$, the optimal bin index decoder
is given by
\begin{equation}
\label{optdec}
w^*(\by)=\mbox{argmax}_{0\le w\le M-1}P(\by|\calC_w)
\end{equation}
where
\begin{equation}
P(\by|\calC_w)\dfn\frac{1}{M_2}\sum_{\bx\in\calC_w}P(\by|\bx)
=\frac{1}{M_2}\sum_{i=wM_2}^{(w+1)M_2-1}P(\by|\bx_i).
\end{equation}
Another, suboptimal decoder, which is natural to consider for bin index
decoding, is the one that
first estimates the index of the transmitted codeword using the ordinary
maximum likelihood (ML) decoder, i.e., 
$\hat{i}_{\mbox{\tiny ML}}(\by)=\mbox{argmax}_{0\le i\le M_1-1}P(\by|\bx_i)$,
and then decodes the bin
index $\hat{w}(\by)$ as the one
that includes that codeword, i.e.,
\begin{equation}
\label{suboptdec}
\hat{w}(\by)=\left\lfloor \frac{\hat{i}_{\mbox{\tiny ML}}(\by)}{M_2}
\right\rfloor.
\end{equation}
In fact, the decoder of the achievability scheme of \cite{Wyner75}
is closer in spirit to 
(\ref{suboptdec}) than to (\ref{optdec}), except that the estimator of $i$ can
even be defined there in terms of joint typicality rather than in terms of ML decoding (in
order to facilitate the analysis). According to the direct part of the coding
theorem in \cite{Wyner75}, for memoryless channels, 
such a decoder is good enough (in spite of its
sub-optimality) for achieving the
maximum achievable information rate, just like decoder (\ref{optdec}).
It therefore seems conceivable that
decoder (\ref{suboptdec}) would achieve the same maximum rate too.
Similar comments apply to the decoder of \cite{GP80}, as well 
as those in many other related works
that involve superposition coding. 

The question that we will address in this paper is what happens if we examine
decoder (\ref{suboptdec}), in comparison to decoder (\ref{optdec}), under the more
refined criterion of the error exponent as a function of the rates $R_1$ and
$R_2$. Would decoder (\ref{suboptdec}) achieve the same optimal error exponent
as the optimal decoder (\ref{optdec})?

By analyzing the exact random coding error exponent associated with
decoder (\ref{optdec}), in comparison to (\ref{suboptdec}), for a given memoryless channel, 
we answer this question affirmatively, at least for
the ensemble of codes described above, where each codeword is selected
independently at random, under the uniform distribution within a given type class.
In particular, our main result is that both decoders achieve the error exponent
given by $E_{\mbox{\tiny r}}(R_1)$, independently
of $R_2$, where $E_{\mbox{\tiny r}}(\cdot)$ is the random
coding error
exponent function of ordinary ML decoding for the above defined
ensemble. In other words, decoder (\ref{suboptdec}) is essentially as good
as the optimal decoder (\ref{optdec}), not only from the viewpoint of
achievable information rates, but moreover, in terms of error exponents. 

The fact that the two decoders have the same error exponent
may appear surprising at first glance. 
It indicates that for a considerable fraction\footnote{Namely, a fraction that
maintains the exponential rate of the probability of the error event.} 
of the error events of (\ref{optdec}), the score $P(\by|\calC_w)$ for a wrong
bin may appear large (enough to exceed the one of the correct bin), mostly
because of an incidental fluctuation in the likelihood of a single codeword
(or a few codewords) within that bin, rather than due to a
collective fluctuation of the entire bin (or a considerable fraction of it).
Thus, it appears conceivable that many of the error events will be
common to both decoders. Now, given that there is a single wrong codeword
(in the entire codebook), 
whose likelihood is exceedingly large, the probability that it would belong
to an incorrect bin is about $(M-1)/M=1-1/M$ (due to symmetry), 
thus roughly speaking, 
erroneous bin index decoding is
essentially as frequent as erroneous decoding of the ordinary ML decoder. Consequently, its
probability depends on $R_2$ so weakly that its asymptotic 
exponent is completely independent of $R_2$.
This independence of $R_2$ means that the
reliability of decoding part of a message ($nR$ out of $nR_1$ nats)
is essentially the same as that of decoding the entire message, no matter
how small or large the size of this partial message may be.

The exponential equivalence of the performance of the two decoders should be
interpreted as an encouraging
message, because the optimal decoder (\ref{optdec}) is extremely difficult to
implement numerically, as the calculation of each score involves the summation of $M_2$ terms
$\{P(\by|\bx_i)\}$, which are typically extremely small numbers for large $n$
(usually obtained from long products of numbers between zero and one).
On the other hand, decoder (\ref{suboptdec}) easily lends itself to calculations
in the logarithmic domain, where products are transformed into
sums, thus avoiding these difficult numerical problems.
Moreover, if the underlying memoryless channel is unknown, decoder
(\ref{suboptdec}) can easily be replaced by a similar decoder that is based on
the universal maximum mutual information (MMI) decoder \cite{CK81}, while it
is less clear how to transform (\ref{optdec}) into a universal decoder.

Yet another advantage of decoder (\ref{suboptdec}) is associated with the
perspective of mismatch. Let
the true underlying channel
$P(\by|\bx_i)$ be replaced by an incorrect assumed channel $P'(\by|\bx_i)$, both in
(\ref{optdec}) and (\ref{suboptdec}). It turns out that the random coding error exponent of the
latter is never worse (and sometimes may be better) than the former. Thus,
decoder (\ref{suboptdec}) is more robust to mismatch.

For the sake of completeness, we also extend our exact error exponent analysis to account
for the hierarchical ensemble of superposition coding (applicable for the
broadcast channel), where first, $M$ cloud
centers, $\bu_0,\bu_1,\ldots,\bu_{M-1}$, are drawn independently at random from a given
type class, and then for each $\bu_w$, $w=0,1,\ldots,M-1$, $M_2$ codewords
$\bx_{wM_2},\bx_{wM_2+1},\ldots,\bx_{(w+1)M_2-1}$ are drawn conditionally
independently from a given conditional type class given $\bu_w$. The resulting
error exponent is the exact\footnote{This is different from earlier work (see
\cite{KM11} and references therein), where lower bounds were derived.}
random coding exponent of the weak decoder in the
degraded broadcast channel model. Here, it is no longer necessarily true that
the error exponent is independent of $R_2$ and that decoders
(\ref{optdec}) and (\ref{suboptdec}) achieve the same exponent. 

Finally, it should be pointed out that in a recent paper \cite{Merhav14},
a complementary study, of the random coding exponent of {\it correct}
decoding, for the optimal bin index
decoder, was carried out for rates above the maximum rate of reliable
communication (i.e., the mutual
information induced by the empirical distribution of the codewords and the
channel). Thus, while this paper is relevant for the legitimate decoder of
the wiretap channel model \cite{Wyner75}, the earlier work \cite{Merhav14} is
relevant for the decoder of the wiretapper of the same model. 

The outline of the remaining part of this paper is as follows. In Section 2,
we establish notation conventions. In Section 3, we formalize the problem
and assert the main theorem concerning the error exponent of
decoders (\ref{optdec}) and (\ref{suboptdec}). Section 4 is devoted to the proof of 
this theorem, and in Section 5, we discuss it. Finally, in Section 6, we
extend our error exponent analysis to the case where the ensemble of random
codes is defined hierarchically.

\section{Notation Conventions}

Throughout the paper, random variables will be denoted by capital
letters, specific values they may take will be denoted by the
corresponding lower case letters, and their alphabets
will be denoted by calligraphic letters. Random
vectors and their realizations will be denoted,
respectively, by capital letters and the corresponding lower case letters,
both in the bold face font. Their alphabets will be superscripted by their
dimensions. For example, the random vector $\bX=(X_1,\ldots,X_n)$, ($n$ --
positive integer) may take a specific vector value $\bx=(x_1,\ldots,x_n)$
in $\calX^n$, the $n$--th order Cartesian power of $\calX$, which is
the alphabet of each component of this vector.
The probability of an event $\calE$ will be denoted by $\mbox{Pr}\{\calE\}$,
and the expectation
operator will be
denoted by
$\bE\{\cdot\}$. 
For two
positive sequences $a_n$ and $b_n$, the notation $a_n\exe b_n$ will
stand for equality in the exponential scale, that is,
$\lim_{n\to\infty}\frac{1}{n}\log \frac{a_n}{b_n}=0$. Thus, $a_n\doteq 0$
means that $a_n$ tends to zero in a super--exponential rate.
Similarly,
$a_n\lexe b_n$ means that
$\limsup_{n\to\infty}\frac{1}{n}\log \frac{a_n}{b_n}\le 0$, and so on.
The indicator function
of an event $\calE$ will be denoted by $\calI\{E\}$. The notation $[x]_+$
will stand for $\max\{0,x\}$.
Logarithms and exponents will be understood to be taken to the natural base
unless specified otherwise.

Probability distributions, associated with sources and channels, will be
denoted by the letters $P$ and $Q$, with subscripts that denote the names of the
random variables involved along with their conditioning, if applicable,
following the customary notation rules in probability theory. For example,
$Q_{XY}$ stands for a generic joint distribution
$\{Q_{XY}(x,y),~x\in\calX,~y\in\calY\}$, $P_{Y|X}$ denotes the
matrix of single--letter transition probabilities 
of the underlying memoryless channel from $X$ to $Y$,
$\{P_{Y|X}(y|x),~x\in\calX,~y\in\calY\}$, and so on. 
Information measures induced
by the generic joint distribution $Q_{XY}$, or $Q$ for short, will be
subscripted by $Q$, for example,
$I_Q(X;Y)$ will denote the corresponding mutual
information, etc. 
The weighted
divergence between two channels, $Q_{Y|X}$ and $P_{Y|X}$, with
weight $P_X$, is defined as
\begin{equation}
D(Q_{Y|X}\|P_{Y|X}|P_X)\dfn\sum_{x\in\calX}P_X(x)\sum_{y\in\calY}Q_{Y|X}(y|x)\ln
\frac{Q_{Y|X}(y|x)}{P_{Y|X}(y|x)}.
\end{equation}
The type class, $\calT(P_X)$, associated with a given empirical probability
distribution $P_X$
of $X$, is the set of all $\bx=(x_1,\ldots,x_n)$, whose empirical distribution
is $P_X$. Similarly, the joint type class of pairs of sequences
$\{(\bu,\bx)\}$ in $\calU^n\times\calX^n$, which is associated with an
empirical joint
distribution $P_{UX}$, will be denoted by $\calT(P_{UX})$. 
Finally, for a given $P_{X|U}$ and $\bu\in\calU^n$, $\calT(P_{X|U}|\bu)$
denotes the conditional type class of $\bx$ given $\bu$ w.r.t.\ $P_{X|U}$,
namely, the set of sequences $\{\bx\}$ whose conditional empirical
distribution w.r.t.\ $\bu$ is given by $P_{X|U}$.

\section{Problem Formulation and Main Result}

Consider a discrete memoryless channel (DMC), defined by a matrix of
single--letter transition probabilities,
$\{P_{Y|X}(y|x),~x\in\calX,~y\in\calY\}$, where $\calX$ and $\calY$ are finite
alphabets. When the channel is fed with an input 
vector $\bx=(x_1,x_2,\ldots,x_n)\in\calX^n$, the output
is a random vector $\bY=(Y_1,\ldots,Y_n)\in\calY^n$, distributed according to
\begin{equation}
P(\by|\bx)=\prod_{t=1}^n P_{Y|X}(y_t|x_t),
\end{equation}
where to avoid cumbersome notation, here and throughout the sequel, 
we omit the subscript ``$\bY|\bX$'' in the
notation of the conditional distribution of the vector channel, from $\calX^n$
to $\calY^n$.
Consider next a codebook, $\calC=\{\bx_0,\bx_1,\ldots,\bx_{M_1-1}\}$, where
$M_1=e^{nR_1}$, and where each $\bx_i$, $i=0,1,\ldots,M_1-1$, is selected
independently at random, under the uniform distribution across the type class
$\calT(P_X)$, where $P_X$ is a given distribution over $\calX$.
Once selected, the codebook $\calC$ is revealed to both the encoder and the
decoder. The codebook $\calC$ is partitioned into $M=e^{nR}$ bins,
$\{\calC_w\}_{w=0}^{M-1}$, each one of size $M_2=e^{nR_2}$ ($R+R_2=R_1$), 
where $\calC_w=\{\bx_{wM_2},\bx_{wM_2+1},\ldots,\bx_{(w+1)M_2-1}\}$.

Let $\bx_I\in\calC$ be transmitted over the channel, where $I$ is a random
variable drawn under the
uniform distribution over $\{0,1,\ldots,M_1-1\}$, independently of the random
selection of the code. 
Let $W=\lfloor I/M_2\rfloor$ designate the random bin index to which $\bx_I$ belongs
and let $\bY\in\calY^n$ be the channel output resulting from the transmission of
$\bx_I$. 

Consider the bin index decoders (\ref{optdec}) and (\ref{suboptdec}), and
define their average error probabilities, as
\begin{equation}
P_e^*=\bE[\mbox{Pr}\{w^*(\bY)\ne W\}],~~~
\hat{P}_e=\bE[\mbox{Pr}\{\hat{w}(\bY)\ne W\}],
\end{equation}
where the probabilities are defined w.r.t.\ the randomness of the index $I$ of
the transmitted codeword (hence the randomness of $W$) and the random operation
of the channel, and the expectations are taken w.r.t.\ the randomness of the
codebook $\calC$.

Our goal is to assess the exact exponential rates of $P_e^*$ and $\hat{P}_e$,
as functions of $R_1$ and $R_2$,
that is,
\begin{equation}
\label{Es}
E^*(R_1,R_2)\dfn \lim_{n\to\infty}\left[-\frac{\ln P_e^*}{n}\right]
\end{equation}
and
\begin{equation}
\label{Eh}
\hat{E}(R_1,R_2)\dfn \lim_{n\to\infty}\left[-\frac{\ln \hat{P}_e}{n}\right].
\end{equation}
At this point, a technical comment is in order. The case $R=0$ ($R_1=R_2$)
should not be understood as a situation where there is only one bin and
$\calC_0=\calC$, since this is a degenerated situation, where there is nothing to
decode as far as bin index decoding is concerned, the probability of error is
trivially zero (just like in ordinary decoding, where there is only one
codeword, which is meaningless). The case $R=0$ should be understood as a case where the number
of bins is at least two, and at most sub-exponential in $n$. On the other
extreme, for $R_2=0$ ($R_1=R$), it is safe to consider each bin as consisting of
a single codeword, rendering the case of ordinary decoding as a special case.

Our main result is the following.
\begin{theorem}
Let $R_1$ and $R_2$ be given ($R_2\le R_1$).
Let $E^*(R_1,R_2)$ and $\hat{E}(R_1,R_2)$ be defined as in eqs.\ (\ref{Es})
and (\ref{Eh}), respectively. Then,
\begin{equation}
E^*(R_1,R_2)=\hat{E}(R_1,R_2)=E_{\mbox{\tiny r}}(R_1)
\end{equation}
where $E_{\mbox{\tiny r}}(R_1)$ is the random coding error exponent function, i.e., 
\begin{equation}
\label{Er}
E_{\mbox{\tiny r}}(R_1)=\min_{Q_{XY}:~Q_X=P_X}\{D(Q_{Y|X}\|P_{Y|X}|P_X)+[I_Q(X;Y)-R_1]_+\}.
\end{equation}
\end{theorem}

\section{Proof of Theorem 1}

For a given $\by\in\calY^n$, and a given joint probability distribution $Q_{XY}$ on
$\calX\times\calY$, let
$N_1(Q_{XY})$ denote the number
of codewords $\{\bx_i\}$ in $\calC_1$ 
whose conditional empirical joint distribution with $\by$ is
$Q_{XY}$, that is
\begin{equation}
N_1(Q_{XY})=\sum_{i=M_2}^{2M_2-1}\calI\{(\bx_i,\by) \in
\calT(Q_{XY})\}.
\end{equation}
We also denote
\begin{equation}
f(Q_{XY})=\frac{1}{n}\ln P(\by|\bx)
=\sum_{(x,y)\in\calX\times\calY}Q_{XY}(x,y)\ln P_{Y|X}(y|x),
\end{equation}
where $Q_{XY}$ is understood to be the joint empirical distribution of
$(\bx,\by)\in\calX^n\times\calY^n$.
Without loss of generality, we assume throughout, that the transmitted codeword
is $\bx_0$, and so, the correct bin is $\calC_0$.
The average probability of error, associated with decoder (\ref{optdec}), is given by
\begin{equation}
\label{startingpoint}
P_e^*\exe \bE\left[\min\{1,M\cdot\mbox{Pr}\{P(\bY|\calC_1)\ge
P(\bY|\calC_0)\}\}\right],
\end{equation}
where the expectation is w.r.t.\ the randomness of $\bX_0,\bX_1,\cdots,\bX_{M_2-1}$ and $\bY$,
and where given $\bX_0=\bx_0$, $\bY$ is
distributed according to 
$P(\cdot|\bx_0)$. 
For a given $\by$, the pairwise error probability,
$\mbox{Pr}\{P(\by|\calC_1)\ge P(\by|\calC_0)\}$, is calculated w.r.t.\
the randomness of $\calC_1=\{\bX_{M_2},\bX_{M_2+1},\ldots,\bX_{2M_2-1}\}$, but for
a given $\calC_0$. To see why (\ref{startingpoint}) is true, first observe
that the right--hand side (r.h.s.) of this equation is simply the (expectation
of the) union bound,
truncated to unity. On the other hand, since the pairwise error events are
conditionally independent given $\calC_0$ and $\by$, the r.h.s.\ times a factor
of $1/2$ (which does not affect the exponent),
serves as a lower bound to the probability of the 
union of the pairwise error events \cite[Lemma A.2]{Shulman03} (see
also \cite[Lemma 1]{SBM07}).

We next move on to the calculation of the pairwise error probability. For a given
$\calC_0$ and $\by$, let 
\begin{equation}
s\dfn \frac{1}{n}\ln\left[\sum_{i=0}^{M_2-1}P(\by|\bx_i)\right],
\end{equation}
and so, the pairwise error probability becomes $\mbox{Pr}\{M_2\cdot P(\by|\calC_1)\ge
e^{ns}\}$, where it is kept in mind that $s$ is a function of
$\calC_0$ and $\by$. Now,
\begin{eqnarray}
\mbox{Pr}\{M_2\cdot P(\by|\calC_1)\ge e^{ns}\}&=&
\mbox{Pr}\left\{\sum_{i=M_2}^{2M_2-1}P(\by|\bX_i)\ge e^{ns}\right\}\\
&=&\mbox{Pr}\left\{\sum_{Q_{X|Y}}N_1(Q_{XY})e^{nf(Q_{XY})}\ge
e^{ns}\right\}\\
&\exe&\mbox{Pr}\left\{\max_{Q_{X|Y}\in\calS(Q_Y)}N_1(Q_{XY})e^{nf(Q_{XY})}\ge
e^{ns}\right\}\\
&=&\mbox{Pr}\bigcup_{Q_{X|Y}\in\calS(Q_Y)}\left\{N_1(Q_{XY})e^{nf(Q_{XY})}\ge
e^{ns}\right\}\\
&\exe&\sum_{Q_{X|Y}\in\calS(Q_Y)}\mbox{Pr}\left\{N_1(Q_{XY})e^{nf(Q_{XY})}\ge
e^{ns}\right\}\\
&\exe&\max_{Q_{X|Y}\in\calS(Q_Y)}\mbox{Pr}\left\{N_1(Q_{XY})\ge
e^{n[s-f(Q_{XY})]}\right\},
\end{eqnarray}
where for a given $Q_Y$, $\calS(Q_Y)$ is defined as the set of all $\{Q_{X|Y}\}$, such that
$\sum_{y\in\calY}Q_Y(y)Q_{X|Y}(x|y)=P_X(x)$ for all $x\in\calX$.
Now, for a given $Q_{XY}$, $N_1(Q_{XY})$ is a binomial random variable with
$e^{nR_2}$ trials and probability of `success' which is of the
exponential order of $e^{-nI_Q(X;Y)}$.
Thus, a standard large deviations analysis (see, e.g., \cite[pp.\ 167--169]{MerhavFnT09}) yields
\begin{equation}
\mbox{Pr}\left\{N_1(Q_{X|Y})\ge
e^{n[s-f(Q_{XY})]}\right\}\exe e^{-nE_0(Q_{XY})},
\end{equation}
where
\begin{eqnarray}
E_0(Q_{XY})&=&\left\{\begin{array}{ll}
[I(Q_{XY})-R_2]_+ & f(Q_{XY})\ge s\\
0 & f(Q_{XY})< s,~f(Q_{XY})\ge s-R_2+I(Q_{XY})\\
\infty & f(Q_{XY})< s,~f(Q_{XY})<
s-R_2+I(Q_{XY})\end{array}\right.\\
&=&\left\{\begin{array}{ll}
I(Q_{XY})-R_2 & f(Q_{XY})\ge s,~R_2<I(Q_{XY})\\
0 & f(Q_{XY})\ge s,~R_2\ge I(Q_{XY})\\
0 & f(Q_{XY})< s,~f(Q_{XY})\ge s-R_2+I(Q_{XY})\\
\infty & f(Q_{XY})< s,~f(Q_{XY})<
s-R_2+I(Q_{XY})\end{array}\right.\\
&=&\left\{\begin{array}{ll}
I(Q_{XY})-R_2 & f(Q_{XY})\ge s,~R_2<I(Q_{XY})\\
0 & f(Q_{XY})\ge s-[R_2-I(Q_{XY})]_+,~R_2\ge I(Q_{XY})\\
\infty & f(Q_{XY})< s-[R_2-I(Q_{XY})]_+\end{array}\right.\\
&=&\left\{\begin{array}{ll}
[I(Q_{XY})-R_2]_+ & f(Q_{XY})\ge s-[R_2-I(Q_{XY})]_+\\
\infty & f(Q_{XY})< s-[R_2-I(Q_{XY})]_+\end{array}\right.
\end{eqnarray}
Therefore, $\max_{Q_{X|Y}\in\calS(Q_Y)}\mbox{Pr}\{N_1(Q_{XY})\ge
e^{n[s-f(Q_{XY})]}\}$ decays according to
$$E_1(s,Q_Y)=\min_{Q_{X|Y}\in\calS(Q_Y)}E_0(Q_{XY}),$$
which is given by
\begin{equation}
E_1(s,Q_Y)=\min\{[I_Q(X;Y)-R_2]_+:~f(Q_{XY})+[R_2-I_Q(X;Y)]_+\ge s\}
\end{equation}
with the understanding that the minimum over an empty set is defined as
infinity. Finally,
\begin{equation}
P_e^*\exe\bE\min\{1,M\cdot e^{-nE_1(S,Q_Y)}\}=\bE e^{-n[E_1(S,Q_Y)-R]_+},
\end{equation}
where the expectation is w.r.t.\ to the randomness of 
\begin{equation}
S=\frac{1}{n}\ln\left[\sum_{i=0}^{M_2-1}P(\bY|\bX_i)\right]
\end{equation}
and the randomness of $Q_Y$, the empirical distribution of $\bY$.
This expectation
will be taken in two steps: first, over the randomness of
$\{\bX_1,\ldots,\bX_{M_2-1}\}$ while $\bX_0=\bx_0$ (the real transmitted codeword)
and $\bY=\by$ are held fixed, and then
over the randomness of $\bX_0$ and $\bY$. Let $\bx_0$ and $\by$ be given
and let $\epsilon>0$ be arbitrarily small. Then,
\begin{eqnarray}
P_e(\bx_0,\by_0) &\dfn&
\bE[\exp\{-n[E_1(S,Q_Y)-R]_+\}|\bX_0=\bx_0,~\bY=\by]\nonumber\\
&=&\sum_s
\mbox{Pr}\{S=s|\bX_0=\bx_0,~\bY=\by\}\cdot
\exp\{-n[E_1(s,Q_Y)-R]_+\}\nonumber\\
&\le&\sum_i \mbox{Pr}\{i\epsilon\le
S<(i+1)\epsilon|\bX_0=\bx_0,~\bY=\by\}\cdot
\exp\{-n[E_1(i\epsilon,Q_Y)-R]_+\},
\end{eqnarray}
where $i$ ranges from $\frac{1}{n\epsilon}\ln P(\by|\bx_0)$ to $R_2/\epsilon$.
Now,
\begin{eqnarray}
\label{ens}
e^{ns}&=&P(\by|\bx_0)+\sum_{i=1}^{M_2-1}P(\by|\bX_i)\nonumber\\
&=& e^{nf(Q_{X_0Y})}+\sum_{Q_{X|Y}}N_0(Q_{XY})e^{nf(Q_{XY})},
\end{eqnarray}
where $Q_{X_0Y}$ is the empirical distribution of $(\bx_0,\by)$ and
$N_0(Q_{XY})$ is the number of codewords in $\calC_0\setminus\{\bx_0\}$ whose
joint empirical distribution
with $\by$ is $Q_{XY}$.
The first term in the second line of (\ref{ens}) 
is fixed at this stage. As for the second term, we have
(similarly as before):
\begin{equation}
\mbox{Pr}\left\{\sum_{Q_{X|Y}}N_0(Q_{XY})e^{nf(Q_{XY})}\ge e^{nt}\right\}\exe
e^{-nE_1(t,Q_Y)}.
\end{equation}
On the other hand,
\begin{equation}
\mbox{Pr}\left\{\sum_{Q_{X|Y}}N_0(Q_{XY})e^{nf(Q_{XY})}\le
e^{nt}\right\}\exe
\mbox{Pr}\bigcap_{Q_{X|Y}}\left\{N_0(Q_{XY})\le e^{n[t-f(Q_{XY})]}\right\}.
\end{equation}
Now, if there exists at least one $Q_{X|Y}\in\calS(Q_Y)$ for which $I_Q(X;Y)< R_2$
and $R_2-I_Q(X;Y)>t-f(Q_{XY})$, then this $Q_{X|Y}$ alone is responsible for a
double exponential decay of the probability of the event $\{N_0(Q_{XY})\le
e^{n[t-f(Q_{XY})]}\}$, let alone the intersection over all
$Q_{X|Y}\in\calS(Q_Y)$.
On the other hand, if for every $Q_{X|Y}\in\calS(Q_Y)$, either $I_Q(X;Y)\ge R_2$
or $R_2-I_Q(X;Y)\le t-f(Q_{XY})$, then we have an intersection of
polynominally many events whose probabilities all tend to unity. Thus, the
probability in question behaves exponentially like an indicator function of
the condition that for every $Q_{X|Y}\in\calS(Q_Y)$, either $I_Q(X;Y)\ge R_2$
or $R_2-I_Q(X;Y)\le t-f(Q_{XY})$, or equivalently,
\begin{equation}
\mbox{Pr}\left\{\sum_{Q_{XY}}N_0(Q_{XY})e^{nf(Q_{XY})}\le
e^{nt}\right\}\exe
\calI\left\{R_2\le\min_{Q_{X|Y}\in\calS(Q_Y)}\{I_Q(X;Y)+[t-f(Q_{XY})]_+\}\right\}.
\end{equation}
Let us now find what is the minimum value of $t$ for which the value of 
this indicator function is unity.
The condition is equivalent to
\begin{equation}
\min_{Q_{X|Y}\in\calS(Q_Y)}\max_{0\le a\le 1}\{I_Q(X;Y)+a[t-f(Q_{XY})]\}\ge R_2
\end{equation}
or, equivalently:
\begin{equation}
\forall Q_{X|Y}\in\calS(Q_Y)~\exists 0\le a\le 1:~~I_Q(X;Y)+a[t-f(Q_{XY})]\ge R_2,
\end{equation}
which can also be written as
\begin{equation}
\forall Q_{X|Y}\in\calS(Q_Y)~\exists 0\le a\le 1:~~t\ge
f(Q_{XY})+\frac{R_2-I_Q(X;Y)}{a}
\end{equation}
or equivalently,
\begin{eqnarray}
t&\ge&\max_{Q_{X|Y}\in\calS(Q_Y)}\min_{0\le a\le
1}\left[f(Q_{XY})+\frac{R_2-I_Q(X;Y)}{a}\right]\\
&=&\max_{Q_{X|Y}\in\calS(Q_Y)}\left[f(Q_{XY})+\left\{\begin{array}{ll}
R_2-I_Q(X;Y) &
R_2\ge I_Q(X;Y)\\ -\infty & R_2< I_Q(X;Y)\end{array}\right.\right]\\
&=&R_2+\max_{\{Q_{X|Y}\in\calS(Q_Y):~I_Q(X;Y)\le R_2\}}[f(Q_{XY})-I_Q(X;Y)]\\
&\dfn&s_0(Q_Y).
\end{eqnarray}
Similarly, it is easy to check that $E_1(t,Q_Y)$ vanishes for $t\le s_0(Q_Y)$.
Thus, in summary, we have
\begin{equation}
\mbox{Pr}\left\{e^{nt}\le \sum_{Q_{X|Y}}N_0(Q_{XY})e^{nf(Q_{XY})}\le
e^{n(t+\epsilon)}\right\}\exe\left\{\begin{array}{ll}
0 & t< s_0(Q_Y)-\epsilon\\
e^{-nE(t,Q_Y)} & t\ge s_0(Q_Y)\end{array}\right.
\end{equation}
Therefore, for a given $(\bx_0,\by)$, the expected error probability w.r.t.\
$\{\bX_1,\ldots,\bX_{M_2-1}\}$ yields
\begin{eqnarray}
P_e(\bx_0,\by)&=&\bE\{e^{-n[E_1(S,Q_Y)-R]_+}|\bX_0=\bx_0,~\bY=\by\}\\
&\le& \sum_{i}\mbox{Pr}\left\{e^{ni\epsilon}\le
\sum_{Q_{X|Y}}N_0(Q_{XY})e^{nf(Q_{XY})}\le
e^{n(i+1)\epsilon)}\right\}\times\nonumber\\
& &\exp(-n[E_1(\max\{i\epsilon,f(Q_{X_0Y})\},Q_Y)-R]_+)\\
&\lexe& \sum_{i\ge s_0(Q_Y)/\epsilon}\exp\{-nE_1(i\epsilon,Q_Y)\}\cdot
\exp(-n[E_1(\max\{i\epsilon,f(Q_{X_0Y})\},Q_Y)-R]_+),
\end{eqnarray}
where the expression $\max\{i\epsilon,f(Q_{X_0Y})\}$ in the argument of
$E_1(\cdot,Q_Y)$ is due to the fact that
\begin{eqnarray}
S&=&\frac{1}{n}\ln\left[e^{nf(Q_{X_0Y})}+\sum_{Q_{X|Y}}N_0(Q_{XY})e^{nf(Q_{XY})}\right]\\
&\ge&\frac{1}{n}\ln\left[e^{nf(Q_{X_0Y})}+e^{ni\epsilon}\right]\\
&\exe&\max\{i\epsilon,f(Q_{X_0Y})\}.
\end{eqnarray}
By using the fact that $\epsilon$ is arbitrarily small, we obtain
\begin{equation}
P_e(\bx_0,\by)\exe \exp(-n[E_1(\max\{s_0(Q_Y),f(Q_{X_0Y})\},Q_Y)-R]_+),
\end{equation}
since the dominant contribution to the sum over $i$ is due to the term
$i=s_0(Q_Y)/\epsilon$ (by the non--increasing monotonicity of the function
$E_1(\cdot,Q_Y)$).
Denoting
$s_1(Q_{X_0Y})=\max\{s_0(Q_Y),f(Q_{X_0Y})\}$,
we then have,
after averaging w.r.t.\ $(\bX_0,\bY)$,
\begin{equation}
E^*(R_1,R_2)=\min_{Q_{Y|X_0}}\{D(Q_{Y|X_0}\|P_{Y|X_0}|P_{X_0})+[E_1(s_1(Q_{X_0Y}),Q_Y)-R]_+\},
\end{equation}
where the random variable $X_0$ is a replica of $X$, that is, $P_{X_0}=P_X$.

We next simplify the formula of $E^*(R_1,R_2)$.
Clearly,
\begin{eqnarray}
E(s_1(Q_{X_0Y}),Q_Y)&=&E_1(\max\{s_0(Q_Y),f(Q_{X_0Y})\},Q_Y)\\
&=&\max\{E_1(s_0(Q_Y),Q_Y),E_1(f(Q_{X_0Y}),Q_Y)\}\\
&=&\max\{0,E_1(f(Q_{X_0Y}),Q_Y)\}\\
&=&E_1(f(Q_{X_0Y}),Q_Y).
\end{eqnarray}
Therefore,
\begin{equation}
E^*(R_1,R_2)=\min_{Q_{Y|X_0}}\{D(Q_{Y|X_0}\|P_{Y|X_0}|P_X)+[E_1(f(Q_{X_0Y}),Q_Y)-R]_+\}.
\end{equation}
Finally, using the simple identity $[[x-a]_+-b]_+\equiv [x-a-b]_+$ ($b\ge 0$), we can
slightly simplify this expression to be
\begin{equation}
E^*(R_1,R_2)=\min_{Q_{Y|X_0}}\{D(Q_{Y|X_0}\|P_{Y|X_0}|P_X)+[I_0(Q_{X_0Y})-R_1]_+\},
\end{equation}
where 
\begin{equation}
I_0(Q_{X_0Y})\dfn\min_{Q_{X|Y}\in\calS(Q_Y)}\{I_Q(X;Y):~f(Q_{XY})+[R_2-I_Q(X;Y)]_+\ge
f(Q_{X_0Y})\}.
\end{equation}
Now, let us define
\begin{equation}
\label{newEr}
E_{\mbox{\tiny r}}'(R_1)\dfn\min_{Q_{Y|X_0}}\{D(Q_{Y|X_0}\|P_{Y|X_0}|P_X)+[I_0'(Q_{X_0Y})-R_1]_+\},
\end{equation}
where 
\begin{equation}
I_0'(Q_{X_0Y})=\min_{Q_{X|Y}\in\calS(Q_Y)}\{I_Q(X;Y):~f(Q_{XY})\ge
f(Q_{X_0Y})\}.
\end{equation}
At this point, $E_{\mbox{\tiny r}}'(R_1)$ 
is readily identified as the ordinary random coding error
exponent associated with ML decoding (i.e., the special case of $E^*(R_1,R_2)$
where $R_2=0$), which is known \cite[p.\ 165, Theorem 5.2]{CK81} to be identical to the
random coding error exponent, $E_{\mbox{\tiny r}}(R_1)$, achieved by maximum mutual
information (MMI) universal decoding, defined similarly, except
that $I_0'(Q_{X_0Y})$ is replaced by
\begin{equation}
I_0''(Q_{X_0Y})=\min_{Q_{X|Y}\in\calS(Q_Y)}\{I_Q(X;Y):~I_Q(X;Y)\ge
I_Q(X_0;Y)\}=I_Q(X_0;Y),
\end{equation}
thus leading to equivalence with eq.\ (\ref{Er}). 

To complete the proof, 
we now argue that $E_{\mbox{\tiny r}}(R_1)=E^*(R_1,R_2)=\hat{E}(R_1,R_2)$. 
The inequality $E_{\mbox{\tiny r}}(R_1)\equiv E_{\mbox{\tiny r}}'(R_1)\ge E^*(R_1,R_2)$ is
obvious since the minimization that defines $I_0'(Q_{X_0Y})$ is over a smaller
set of distributions than the one that defines $I_0(Q_{X_0Y})$. On the other
hand, the converse inequality, $E_{\mbox{\tiny r}}(R_1)\le E^*(R_1,R_2)$, is also true 
because of the following consideration. We claim that
\begin{equation}
E_{\mbox{\tiny r}}(R_1)\le E'(R_1,R_2)\le \hat{E}(R_1,R_2)\le E^*(R_1,R_2),
\end{equation}
where definitions and explanations are now in order:
As defined, $E_{\mbox{\tiny r}}(R_1)$ is the random coding error exponent associated with
ordinary ML decoding and the ordinary probability of error for a random code
at rate $R_1$. Now, let $E'(R_1,R_2)$ be defined as
the random coding exponent of the ML decoder, where only errors associated with
winning codewords that are outside the correct bin $\calC_0$ are counted. In
other words, assuming that $\bx_0$ was transmitted, this is the exponent of
the probability of the event $\{\max_{i\ge M_2}P(\by|\bx_i)\ge P(\by|\bx_0)\}$.
Since this error event is a subset of the ordinary error event, its
exponent is at least as large as $E_r(R_1)$, hence the first inequality. Now,
$\hat{E}(R_1,R_2)$, which is the error exponent of decoder (\ref{suboptdec}), is in fact,
the exponent of the probability of the event $\{\max_{i\ge M_2}P(\by|\bx_i)\ge
\max_{i<M_2}P(\by|\bx_i)\}$ (given $\bx_0)$), which in turn is a subset of the previous
error event defined, hence the second inequality. Finally, the last inequality
follows from the optimality of decoder (\ref{optdec}), whose error exponent
cannot be smaller than that of (\ref{suboptdec}). Thus, we conclude that
all inequalities are, in fact, equalities, and so,
$E^*(R_1,R_2)=\hat{E}(R_1,R_2)=E_{\mbox{\tiny r}}(R_1)$, completing the
proof of Theorem 1.

\section{Discussion}

When $R_2=0$, that is, a subexponential number of codewords within each bin,
Theorem 1 is actually
not surprising since $\sum_{\bx\in\calC_w}P(\by|\bx)\exe
\max_{\bx\in\calC_w}P(\by|\bx)$, but for $R_2> 0$, the results are not quite
trivial (at least for the author of this article).
As explained in the Introduction, the intuition is that the error
probability is dominated by few codewords within some bin, whose likelihood
score is exceptionally high. Note also that bin index decoding is different from 
the situation in list decoding,
where even for a subexponential list size, the error exponent is improved.
This is not surprising, because in list decoding, the list depends on
the likelihood scores, and they are not given by a fixed bin, which is
arbitrary.

Theorem 1 tells us that, under the
ordinary random coding regime, decoding only a part of a message (say, a
header of $nR$ nats out of the total of $nR_1$)
is as reliable as decoding the entire message, as far as error exponents go.
As discussed in the Introduction, decoder (\ref{suboptdec}) is easier to
implement. It is also clear how to universalize this decoder: for an unknown DMC, replace
$\hat{i}_{\mbox{\tiny ML}}(\by)$ in (\ref{suboptdec}) by
\begin{equation}
\hat{i}_{\mbox{\tiny MMI}}(\by)= \mbox{argmax}_{0\le i\le M_1-1}I_Q(X_i;Y),
\end{equation}
where $I_Q(X_i;Y)$ designates the empirical mutual information induced by
$(\bx_i,\by)$. This universal bin index decoder still achieves $E_r(R_1)$.

As for the mismatched case, the only change in the derivation in Section 4 
is that the definition of the
function $f(Q_{XY})$ is changed to $f(Q_{XY})=\sum_{x,y}Q(x,y)\ln P_{Y|X}'(y|x)$
(or more generally. to an arbitrary function of $Q_{XY}$), where $P_{Y|X}'(y|x)$ is
the mismatched channel. Here, it will still be true that $E_r'(R_1)$ defined as
in (\ref{newEr}) (but with $f$ being
redefined) is not smaller than the corresponding $E^*(R_1,R_2)$, but the converse inequality
(that was leading to equality before) can no longer be claimed since it was
based on the optimality of decoder (\ref{optdec}), but now both decoders are suboptimal.
This means that, for the purpose of bin index decoding, decoder
(\ref{suboptdec}), but with $P_{Y|X}$ replaced by $P_{Y|X}'$,
is never worse than the corresponding mismatched version of decoder (\ref{optdec}).

\section{Extension to Hierarchical Ensembles}

Consider again a random code $\calC$ of size $M_1=e^{nR_1}$, but this time, it
is drawn from a different ensemble, which is in the spirit of the ensemble of the direct part
of the coding theorem for the degraded broadcast channel (see, e.g.,
\cite[Section 15.6.2]{CT06}). Specifically, 
let $\calU$ be a finite alphabet, let $P_U$ be a given probability
distribution on $\calU$, and let $P_{X|U}$ be a given matrix of conditional
probabilities of $X$ given $U$.
We first select, independently at random, $M=e^{nR}$ $n$--vectors (``cloud
centers''), $\bu_0,\bu_1,\ldots,\bu_{M-1}$, all under the uniform distribution
over the type class $\calT(P_U)$. Next, for each $w=0,1,\ldots,M-1$, we select
conditionally independently (given $\bu_w$), $M_2=e^{nR_2}$ codewords,
$\bx_{wM_2},\bx_{wM_2+1},\ldots,\bx_{(w+1)M_2-1}$, under the uniform
distribution across the conditional type class $\calT(P_{X|U}|\bu_w)$.
Obviously, the ensemble considered in the previous
sections is a special case, where $P_U$ is a degenerate distribution, putting
all its mass on one (arbitrary) letter of $\calU$. All other quantities are
defined similarly as before.

We next present a more general formula of $E^*(R_1,R_2)$, the exact random
coding error exponent
of decoder (\ref{optdec}), that accommodates the above defined
ensemble. This is then the exact random coding error exponent of 
the optimal decoder for the weak user in the degraded broadcast channel.
Here, we no longer claim that $E^*(R_1,R_2)$ is independent of $R_2$
and that it is achieved by decoder (\ref{suboptdec}) as well.

To present the formula of $E^*(R_1,R_2)$, we first need a few definitions.
For a given generic joint distribution $Q_{UXY}$, of the random variables
$U$, $X$, and $Y$, let $I_Q(X;Y|U)$ denote the conditional mutual
information between $X$ and $Y$ given $U$. For a given marginal $Q_{UY}$ of $(U,Y)$,
let $\calS(Q_{UY})$ denote the set of conditional distributions $\{Q_{X|UY}\}$
such that $\sum_{y}Q_{UY}(u,y)Q_{X|UY}(x|u,y)=P_{UX}(u,x)$ for every
$(u,x)\in\calU\times\calX$, where $P_{UX}=P_U\times P_{X|U}$.
We first define
\begin{equation}
\label{E1}
E_1(s,Q_{UY})=\min_{Q_{X|UY}\in\calS(Q_{UY})}
\{[I_Q(X;Y|U)-R_2]_+:~f(Q_{XY})+[R_2-I_Q(X;Y|U)]_+\ge s\},
\end{equation}
where $s$ is an arbitrary real number.
Next, for a given marginal $Q_Y$, define
\begin{equation}
\label{E2}
E_2(s,Q_Y)=\min_{Q_{U|Y}}[I_Q(U;Y)+E_1(s,Q_{UY})],
\end{equation}
where the minimization is across all $\{Q_{U|Y}\}$ such that
$\sum_{y}Q_{Y}(y)Q_{U|Y}(u|y)=P_U(u)$ for every $u\in\calU$.
Finally, let
\begin{equation}
\label{s0}
s_0(Q_{U_0Y})=R_2+\max_{\{Q_{X|U_0Y}\in\calS(Q_{U_0Y}):~I_Q(X;Y|U_0)\le
R_2\}}[f(Q_{XY})-I_Q(X;Y|U_0)],
\end{equation}
and 
\begin{equation}
\label{s1}
s_1(Q_{U_0X_0Y})=\max\{s_0(Q_{U_0Y}),f(Q_{X_0Y})\}.
\end{equation}
Our extended formula for $E^*(R_1,R_2)$ is given in the following theorem.

\begin{theorem}
Let $R_1$ and $R_2$ be given ($R_2\le R_1$) and let the ensemble of codes be
defined as in the first paragraph of this section. Then,
\begin{equation}
E^*(R_1,R_2)=\min_{Q_{Y|X_0U_0}}\{D(Q_{Y|X_0U_0}\|P_{Y|X_0}|P_{U_0X_0})+[E_2(s_1(Q_{U_0X_0Y}),Q_Y)-R]_+\}.
\end{equation}
where $(U_0,X_0)$ is a replica of $(U,X)$, i.e., $P_{U_0X_0}=P_{UX}$.
\end{theorem}

\vspace{0.2cm}

\noindent
{\it Proof Outline.}
The proof of Theorem 2 is quite a straightforward generalization of the proof
of Theorem 1, which was given in full detail in Section 4. We will therefore
give here merely an outline with highlights mostly on the differences.
Once again, we start from the expression,
\begin{eqnarray}
P_e&\exe& \bE\left[\min\{1,M\cdot\mbox{Pr}\{P(\bY|\calC_1)\ge
P(\bY|\calC_0)\}\}\right],
\end{eqnarray}
where this time, the expectation is w.r.t.\ the randomness of $\bU_0$,
$\calC_0$ and $\bY$, with the latter being the channel output in response to
the input $\bX_0$ (which is again, the transmitted codeword, without loss of generality). 
Here, for a given $\by$, the pairwise error probability,
$\mbox{Pr}\{P(\by|\calC_1)\ge P(\bY|\calC_0)\}$, is calculated w.r.t.\
the randomness of $\bU_1$, $\calC_1=\{\bX_{M_2},\bX_{M_2+1},\ldots,\bX_{2M_2-1}\}$, but for
a given $\bu_0$, and $\calC_0$.\footnote{The reason that the randomness of $\bU_1$ is
accommodated already at this stage, is that this way, the $M-1$ pairwise error events
are all independent (given $\bU_0$, $\calC_0$ and $\by$) and have identical
probabilities, and so, the truncated union bound,
$\min\{1,M\cdot\mbox{Pr}\{P(\by|\calC_1)\ge P(\by|\calC_0)\}\}$, remains exponentially tight, as before.}

Defining $s$ as in the proof of Theorem 1, the pairwise error probability is
calculated once again, using the large deviations properties of $N_1(\cdot)$,
which are now binomial random variables given $\bu_1$. Thus, we first
calculate the pairwise error probability conditioned on $\bU_1=\bu_1$, and
then average over $\bU_1$.
Now, for a given $Q_{UXY}$, designating the joint empirical distribution of a randomly
chosen $\bx$ together with $(\bu_1,\by)$, the binomial random variable
$N_1(Q_{UXY})$ has $e^{nR_2}$ trials and probability of success which is of
the exponential order of $e^{-nI_Q(X;Y|U)}$. Everything else in this large
deviations analysis remains intact.
Thus, $E_0(Q_{XY})$, in the proof
of Theorem 1, should be replaced by $E_0(Q_{UXY})$,
which is defined by
\begin{equation}
E_0(Q_{UXY})=\left\{\begin{array}{ll}
[I_Q(X;Y|U)-R_2]_+ & f(Q_{XY})\ge s-[R_2-I_Q(X;Y|U)]_+\\
\infty & f(Q_{XY})< s-[R_2-I_Q(X;Y|U)]_+\end{array}\right.
\end{equation}
Therefore, $E_1(s,Q_{UY})$ of 
the proof of Theorem 1, should now be replaced by
$E_1(s,Q_{UY})$ as defined eq.\ (\ref{E1}).
The conditional pairwise error probability, that includes also conditioning
on $\bU_1=\bu_1$, is then of the exponential order of $e^{-nE_1(s,Q_{UY})}$.
After averaging this exponential function w.r.t.\ the randomness of $\bU_1$
(thus relaxing the conditioning $\bU_1=\bu_1$), the resulting expression
becomes of the exponential order of $e^{-nE_2(s,Q_Y)}$, where $E_2(s,Q_Y)$ is
defined as in (\ref{E2}). The remaining part of the proof is exactly in the footsteps
of the proof of Theorem 1, except that here, the simplifications given near
the end of the proof do not seem to hold anymore. $\Box$

\end{document}